\documentclass[10pt,letterpaper,twocolumn]{article} %% two column, final layout

\usepackage{ol2}
\usepackage[draft]{hyperref}
\usepackage{amsmath}

\begin{document}

\twocolumn[ %% activate for two-column option

\title{$\mathcal{PT}$-symmetric mode-locking}

%% For REVTeX it is possible to automate superscript and e-mail callouts with the superscriptaddress option; see REVTeX4 documentation.

\author{Stefano Longhi}

\address{Dipartimento di Fisica, Politecnico di Milano and Istituto di Fotonica e Nanotecnologie del Consiglio Nazionale delle Ricerche, Piazza L. da Vinci 32, I-20133 Milano, Italy (stefano.longhi@polimi.it)}

\begin{abstract}
Parity-time ($\mathcal{PT}$) symmetry is one of the most important accomplishments in optics over the past decade. Here 
the concept of $\mathcal{PT}$ mode-locking of a laser is introduced, in which active phase locking of cavity axial modes is realized by  {\it asymmetric} mode coupling in a complex time crystal. $\mathcal{PT}$ mode-locking shows a transition from single to double pulse emission as the $\mathcal{PT}$ symmetry breaking point is crossed. The transition can show a turbulent behavior, depending on a dimensionless modulation parameter that plays the same role as the Reynolds number in hydrodynamic flows. 
\end{abstract}

% 270.3430   Laser theory
% 190.3100   Instabilities and chaos
% 140.3518   Lasers, frequency modulated 
% 000.1600   Classical and quantum physics
 ] %% activate for two-column option

Mode-locking (ML) of a laser is a rather complex phenomenon in which many cavity axial modes lock together to generate ultrashort pulses \cite{r1,r2}. ML has been a milestone of laser science, with major applications to such different areas as ultrafast spectroscopy, high-speed optical communications, metrology, attosecond science, etc. Traditionally, ML methods are classified into active and passive methods \cite{r1,r2}. In active ML phase locking of cavity axial modes is forced by either intracavity amplitude (AM) or frequency (FM) modulation, which provides a {\it symmetric} transfer of the optical power among the longitudinal cavity modes \cite{r1}. The hindered complex dynamics of active and passive ML has attracted great interest since the invention of lasers \cite{r2}, providing an experimentally accessible laboratory tool for the investigation of universal phenomena of dissipative dynamical systems and phase transitions \cite{r2bis,r3,r4,r5,r6,r7,r8,r9,r10,r11,r13,r16,r17}. For example, actively ML lasers can show excess noise and turbulent behavior similar to hydrodynamics flows \cite{r2bis,r3,r4,r8,r10}, or a light-mode transition similar to Bose-Einstein condensation \cite{r11}. Phenomena like Anderson localization, Bloch oscillations and metal-insulator phase transitions typical of the solid-state physics can be observed in the spectrum of ML lasers \cite{r5,r7,r9}. Even the emergence of the mode-locked state from initial noise is an intrinsically singular transition, which has been measured in a recent experiment \cite{r17}. 
%Phase locking is not restricted to conventional lasers, rather it can be found in exotic systems like random lasers \cite{r13}. \\   
Recently, an interesting link has been established between phase transitions in certain dissipative systems driven out of equilibrium and parity-time ($\mathcal{PT}$) symmetric models \cite{r18,r19}. $\mathcal{PT}$-symmetry, originally introduced in quantum physics as a complex extension of quantum mechanics \cite{r20}, has provided a fruitful concept in optics in the past few years (see, for instance, \cite{r21,r22,r23,r24,r25,r26,r27,r27bis} and references therein). $\mathcal{PT}$ optical structures show balanced gain and loss distributions, undergoing a symmetry breaking phase transition when the gain/loss contrast is increased. A particular and important class of $\mathcal{PT}$-symmetric systems is provided by periodic optical media, so-called complex crystals \cite{r27,r28,r29}, which  are one-way invisible at the symmetry breaking point \cite{r27,r29,r30}.\par
In this Letter the concept of $\mathcal{PT}$-symmetric ML is introduced, in which the symmetry breaking phase transition can show a turbulent behavior of laser pulse emission. As compared to conventional active ML, in the $\mathcal{PT}$-symmetric ML  transfer of the optical power between adjacent cavity axial modes is {\it asymmetric}. The asymmetry of mode coupling is realized by a suitable combination of intracavity AM and FM, so that the optical pulse circulating in the cavity is repeatedly scattered off by a {\it complex time crystal} \cite{r31}. Assuming exact synchronism between AM/FM modulation frequency $\omega_m$ and cavity axial mode separation $\omega_{ax}$, i.e. $\omega_m=\omega_{ax}$, the equation of motion for the pulse envelope $\psi(t,n)$ circulating in the cavity at successive round trips is given by the ML master equation \cite{r2,r3,r7,r32,r33}
\begin{equation}
\frac{\partial \psi}{\partial n}  =  \left( g-l+\mathcal{D}_g \frac{\partial^2}{\partial t^2}- \Delta_{AM} (1-\cos(\omega_m t)) +i \Delta_{FM} \sin (\omega_m t) \right) \psi 
\end{equation}
where $t$ is the fast time variable that varies over the cavity round trip interval ($-T_m/2<t<T_m/2$), $n$ is the round-trip number, $\omega_m= 2 \pi / T_m= \omega_{ax}$ is the modulation frequency, $g=g(n)$ and $l$ are the saturated gain and loss per transit in the cavity, $\mathcal{D}_{g}= 1 / \omega_g^2$ is the spectral filtering parameter determined by gain bandwidth $\omega_g$ of the cavity, and $\Delta_{AM}$, $\Delta_{FM}$ are the modulation depths of the amplitude and phase modulators, respectively. For a slow-gain medium, saturated gain $g$ obeys the rate equation
\begin{equation}
\frac{dg}{dn}= -\gamma_{\|} (g-g_0+g \mathcal{P})
\end{equation}
where $g_0$ is the small-signal gain from pumping, $\gamma_{\|}=T_m / \tau$ is the ratio between cavity transit time $T_m$
and upper laser level lifetime $\tau$ ($\tau \gg T_m$), and $\mathcal{P}=(1/T_m) \int_{-T_m/2}^{T_m/2} dt |\psi(t,n)|^2$  is the average laser power
normalized to the saturation power. As compared to the conventional AM ML regime, which is attained from Eq.(1) by assuming $\Delta_{FM}=0$, the addition of the quarter-phase-shifted FM modulation yields asymmetric power transfer between adjacent axial modes of the cavity. This can be seen by writing Eq.(1) in the frequency (spectral) domain. After setting $\psi(t,n)=\sum_{q} \phi_q (n) \exp(i q \omega_m t)$, the following coupled equations for the spectral mode amplitudes $\phi_q(n)$ are obtained
\begin{equation}
\frac{d \phi_q}{dn}=(g-l-\Delta_{AM}-\mathcal{D}_g \omega_m^2 q^2) \phi_q+\Delta_{-} \phi_{q+1}+\Delta_{+} \phi_{q-1},
\end{equation}
  \begin{figure}[htbp]
\centerline{\includegraphics[width=8.4cm]{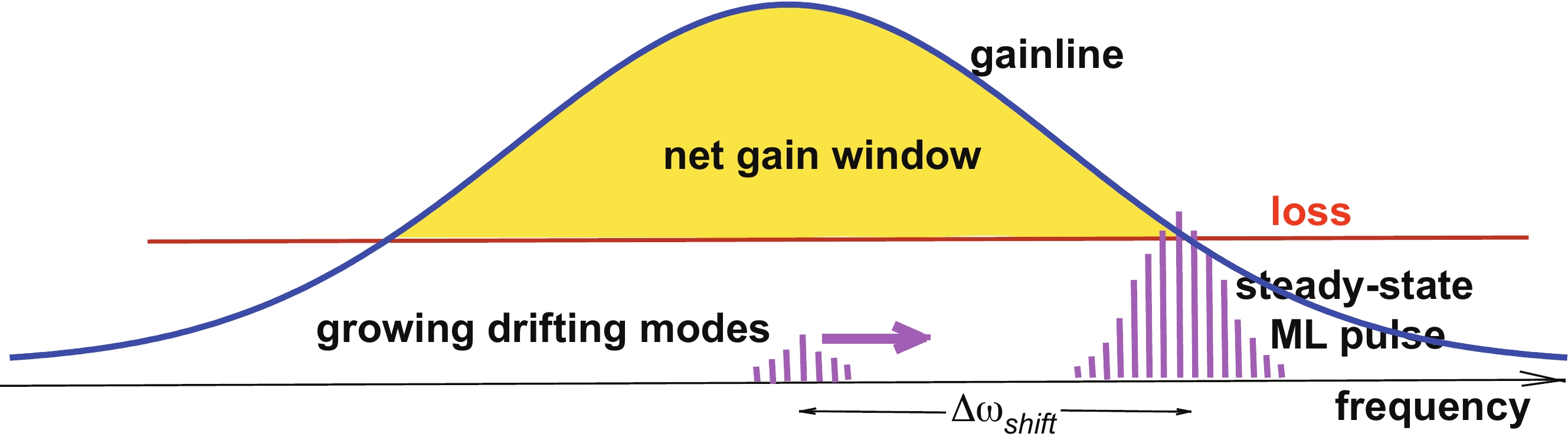}} \caption{ \small
(Color online) Schematic of transient amplification of spectral modes in the $\mathcal{PT}$-symmetric ML. Owing to the frequency modulation, the steady-state ML pulse spectrum is shifted by $\Delta \omega_{shift}$ away from the center of the gainline, thus requiring an extra gain. As a result, a net spectral gain window (shaded region) arises near the center of the gainline. Small-amplitude cavity axial modes near the center of the gainline can be thus transiently amplified while advected  away.}
\end{figure} 
where $q=0, \pm1, \pm2,...$ is the axial mode number and where we have set $\Delta_{\pm}=(\Delta_{AM} \pm \Delta_{FM})/2$. Note that when the FM modulator is switched off ($\Delta_{FM}=0$), i.e. in ordinary AM ML, mode coupling is symmetric $\Delta_+=\Delta_-$, while it becomes asymmetric when the FM modulator is switched on, with unidirectional coupling  ($\Delta_-=0$) at $\Delta_{FM}=\Delta_{AM}$. Such an asymmetric mode coupling is a rather general feature of complex crystals \cite{r34}. Interestingly, the pulse dynamics at successive transits in the cavity as described by Eq.(1) realizes a Wick-rotated $\mathcal{PT}$-symmetric model \cite{r19}, i.e. it can be formally written as $i \partial_{\tau} \psi= \hat{H} \psi$ where $\tau=-in$ is complex (Wick-rotated) time \cite{r19} and $\hat{H}$ is the $\mathcal{PT}$-symmetric Hamiltonian with complex sinusoidal potential
\begin{equation}
\hat{H} \equiv-\mathcal{D}_g \partial^2_t- \Delta_{AM} \cos(\omega_m t)-i \Delta_{FM} \sin (\omega_m t) +l+\Delta_{AM}-g
\end{equation}
  which depends parametrically on the saturated gain $g$. As is well known, assuming $g=l+\Delta_{AM}$ the energy spectrum of $\hat{H}$ is entirely real for $\Delta_{FM}<\Delta_{AM}$ (unbroken $\mathcal{PT}$ phase), whereas it is formed by complex-conjugate pairs for $\Delta_{FM}>\Delta_{AM}$ (broken $\mathcal{PT}$ phase) \cite{r23,r24}. Owing to Wick rotation of time, in the physical problem Eq.(1)  $\mathcal{PT}$ symmetry breaking of $\hat{H}$ corresponds to a transition from a regime of a simple lowest-threshold ML state, i.e. a single ML pulse, to a regime of doubly-degenerate lowest-threshold ML state, i.e. to ML pulse doubling \cite{r19,r33}. 
Such a transition has a simple physical explanation in terms of ordinary Kuizenga-Siegman theory of AM and FM laser ML \cite{r1,r35}: in the limit $\Delta_{FM} \ll \Delta_{AM}$ the FM signal can be regarded as a small perturbation, and thus the laser operates in the AM ML  emitting a single Gaussian ML pulse centered at $t=0$. On the other hand, in the opposite limit $\Delta_{FM} \gg \Delta_{AM}$ the AM signal can be regarded as a small perturbation, and the laser operates this time in the so-called FM ML regime, which is known to sustain two threshold-degenerate ML pulses  centered at $t= \pm T_m/4$ \cite{r35}. Pulse doubling transition was previously investigated in  Ref.\cite{r33}, however it was not related to a $\mathcal{PT}$ symmetry breaking phase transition.  Here we wish to show that, owing to the non-normal nature of $\hat{H}$ \cite{r3,r4,r18}, the $\mathcal{PT}$ phase transition can show a turbulent behavior. To this aim, let us assume $\Delta_{FM} \leq  \Delta_{AM}$ (unbroken $\mathcal{PT}$ phase), so that there is a single ML pulse centered at around $t \simeq 0$. In such a regime, the pulse dynamics can be captured within a parabolic approximation of the complex sinusoidal potential near the minimum $t=0$ of AM loss, i.e. by letting $\cos(\omega_mt) \simeq 1-\omega_m^2 t^2/2$ and $\sin (\omega_m t) \simeq \omega_m t$ in Eqs.(1) and (4). This yields the effective $\mathcal{PT}$-symmetric Hamiltonian
  \begin{equation}
  \hat{H}_{eff}=-\mathcal{D}_g \frac{\partial^2 }{\partial t^2}+ \frac{1}{2} \Delta_{AM} \omega_m^2 (t-i \delta)^2+\sigma
  \end{equation}
    \begin{figure}[htbp]
\centerline{\includegraphics[width=\linewidth]{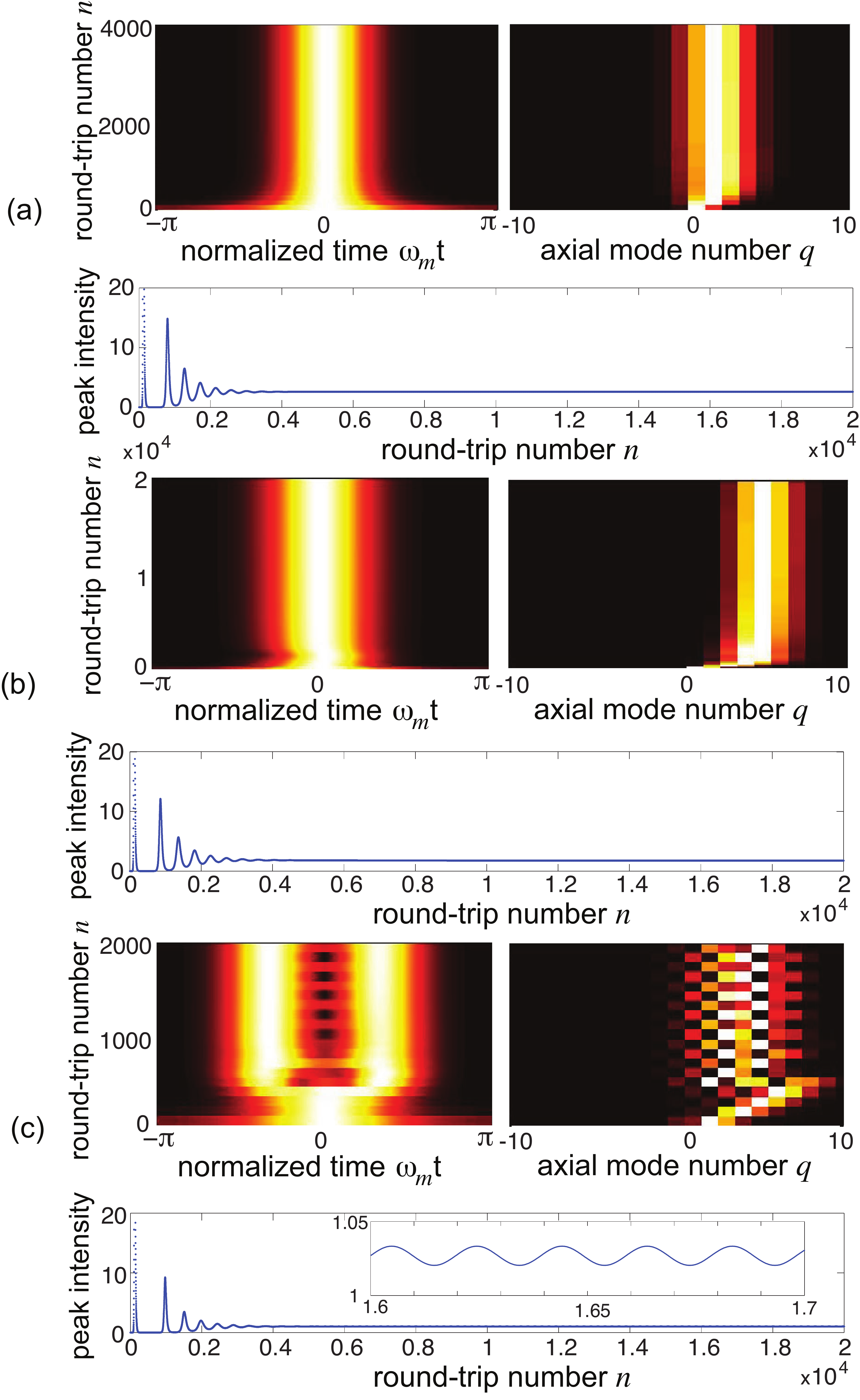}} \caption{ \small
(Color online) Transient formation of laser ML. The figures show the numerically-computed evolution of normalized pulse intensity (left panel), pulse spectrum (right panel) and peak pulse intensity (bottom panel) at successive round trips, starting form a small random field, for increasing values of the FM parameter $\Delta_{FM}$: (a) $\Delta_{FM}=0.005$ (unbroken $\mathcal{PT}$ phase),  (b) $\Delta_{FM}=0.01$ (symmetry breaking point), and (c) $\Delta_{FM}=0.02$ (broken $\mathcal{PT}$ phase). Other parameter values are: $\Delta_{AM}=0.01$, $\omega_g / \omega_m=50$, $\gamma_{\|}=1 \times 10^{-3}$, $l=0.04$, and $g_0=0.15$. The inset in the bottom panel of (c) shows an enlargement of the peak pulse intensity evolution after transient relaxation oscillations, showing small undamped oscillations.}
\end{figure} 
where we have set $\sigma \equiv l-g+ \Delta_{FM}^{2} /(2 \Delta_{AM})$ and $\delta \equiv \Delta_{FM}/(\omega_m \Delta_{AM})$. Interestingly, $\hat{H}_{eff}$ describes the Hamiltonian of  a $\mathcal{PT}$-symmetric quantum harmonic oscillator \cite{ruff}, which is obtained from the ordinary (Hermitian) quantum oscillator Hamiltonian  $\hat{H}_{QO}=-\mathcal{D}_g \partial^2_x +(1/2) \Delta_{AM} \omega_m^2 x ^2+\sigma$ after complexification of the spatial variable $x=t-i \delta$. $\hat{H}_{eff}$ and $\hat{H}_{QO}$ thus shear the same eigenvalues, whereas the eigenmodes of $\hat{H}_{eff}$ are obtained from the Gauss-Hermite modes of $\hat{H}_{QO}$  after the substituion $x =t- i \delta$. From the eigenvalues of $\hat{H}_{QO}$ one can then readily obtain the gain thresholds $g_{0 \; th}$ of the various Gauss-Hermite modes. In particular, the lowest-threshold mode is the fundamental Gaussian state, given by
\begin{equation}
\psi_0(t)=\exp[- \rho(t-i \delta)^2]
\end{equation}
with corresponding gain threshold 
\begin{equation}
g_{0 \; th}=l+\sqrt{\frac{\Delta_{AM} \mathcal{D}_g \omega_m^2}{2}} + \frac{\Delta_{FM}^{2}}{2 \Delta_{AM}},
\end{equation}
 where we have set $\rho \equiv  (\omega_m/2) (\Delta_{AM}/2 \mathcal{D}_g)^{1/2}$.
 \begin{figure}[htbp]
\centerline{\includegraphics[width=\linewidth]{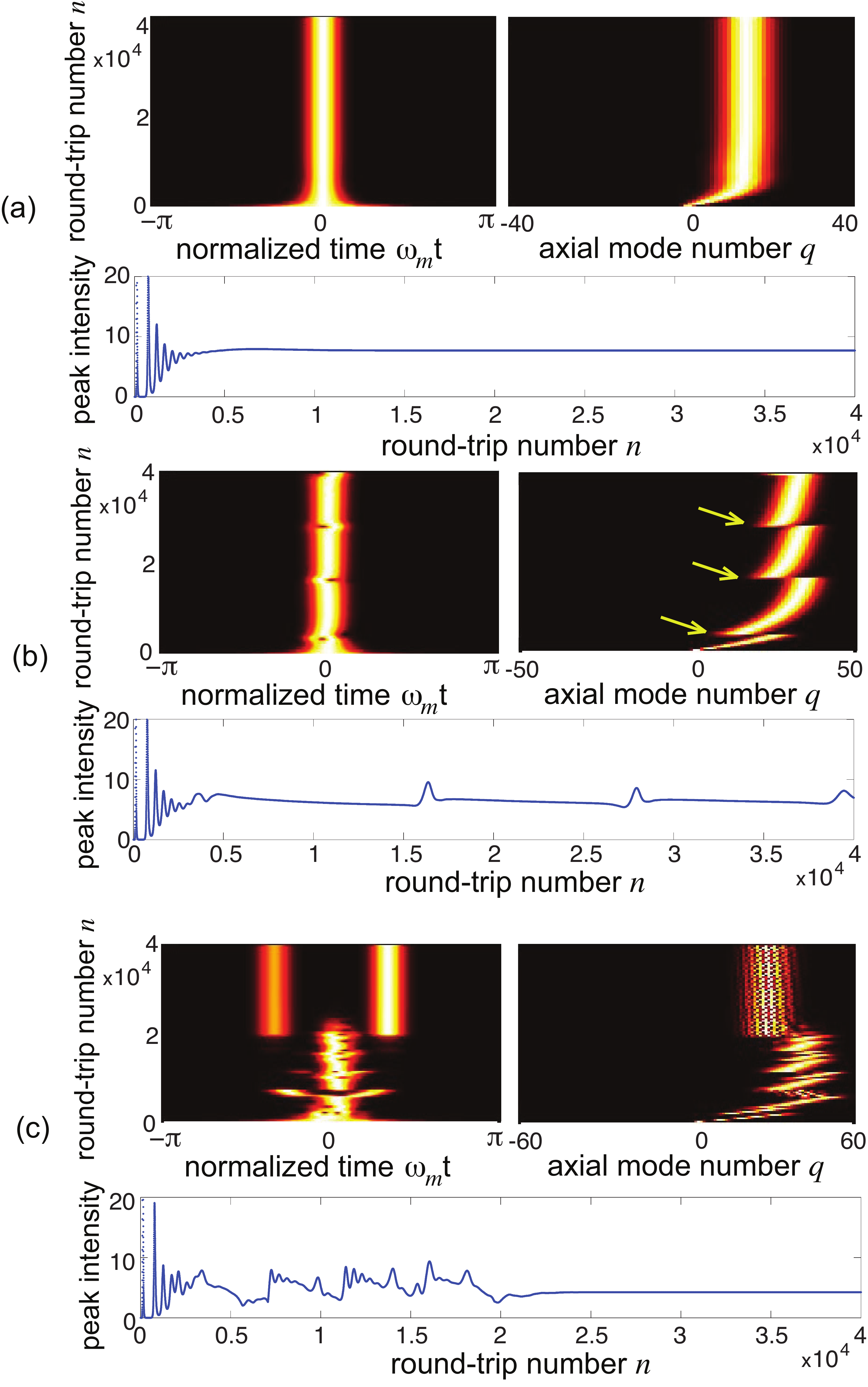}} \caption{ \small
(Color online) Same as Fig.2, except for $\omega_g / \omega_m=400$.}
\end{figure} 
Note that, like in the ordinary AM ML, the fundamental Gaussian state is centered at $t=0$, i.e. at the minimum of AM modulation loss, however the effect of a non-vanishing $\delta$ is to {\it spectrally shift} the ML pulse away from the center of the gainline by the amount 
\begin{equation}
\Delta \omega_{shift}=2 \rho \delta=\Delta_{FM} (2 \Delta_{AM} \mathcal{D}_g)^{-1/2} ,
\end{equation}
 as one can see from Eq.(6). Moreover, the fundamental Gaussian state turns out to be linearly stable: it saturates the gain $g$ so as all other higher-order modes experience net loss. However, owing to the complex displacement $\delta \neq 0$, $\hat{H}_{eff}$ is a non-normal operator, i.e. $\hat{H}_{eff}$ does not commute with its Hermitian adjoint, its eigenmodes are not orthogonal, and transient amplification of small perturbations can be thus observed \cite{r3, r4,r10,r36}. The maximum energy amplification $G$ of perturbations is given by the Petermann excess noise factor \cite{r10}, which is given by $G=\langle \psi_0, \psi_0 \rangle   \langle \psi^{\dag}_0, \psi^{\dag}_0 \rangle /  | \langle \psi_0, \psi^{\dag}_0 \rangle|^2$, where $\langle f , g \rangle = \int_{-\infty}^{\infty} dt f^*(t) g(t) $ denotes the usual (Hermitian) scalar product and $\psi_0^{\dag}(t)$ is the adjoint mode of $\psi_0(t)$, which is simply obtained from $\psi_0(t)$  after the change $\delta \rightarrow -\delta$ on the right hand side of Eq.(6). One obtains $G=\exp(2R)$, where we have set
\begin{equation}
R \equiv 2 \rho \delta^2= (\omega_g / \omega_m) \Delta_{FM}^{2} (2 \Delta_{AM}^3)^{-1/2}  % \frac{\omega_g}{\omega_m} \frac{\Delta_{FM}^{2}}{\sqrt {2 \Delta_{AM}^{3}}}.
\end{equation}
 From a physical viewpoint, the transient amplification of perturbations can be explained by the appearance of a net {\it spectral} gain window, as schematically shown in Fig.1.  In fact, within the  parabolic approximation of the complex sinusoidal $\mathcal{PT}$ potential, the dynamics of the spectral modes, as given by Eq.(3), can be cast in the form
\begin{equation}
\frac{\partial \phi}{\partial n}= (g-l) \phi+\frac{1}{2} \Delta_{AM} \frac{\partial^2 \phi}{\partial q^2}-\mathcal{D}_g \omega_m^2 q^2 \phi-\Delta_{FM} \frac{\partial \phi}{\partial q},
\end{equation}
where $\phi(q,n)$ is the spectrum (Fourier transform) of the pulse envelope at the $n$-th round trip in the cavity. Equation (10) is readily obtained from Eq.(3) by considering the mode index $q$ as a continuous variable, $ \phi_q(n) \rightarrow \phi(q,n)$, and after setting $\phi_{q \pm 1} \simeq  \phi(q) \pm ( \partial \phi / \partial  q) +(1/2) (\partial^2 \phi / \partial q^2)$. When $\Delta_{FM}=0$, the pulse spectrum is Gaussian and centered at $q=0$, i.e. at the center of the gainline. However, for asymmetric mode coupling $\Delta_{FM} \neq 0$ a drift term arises in Eq.(10) (the last term on the right hand side), which shifts the pulse spectrum away from the center of the gainline by the amount $\Delta \omega_{shift}$ [Eq.(8)]. Such a spectral shift leads to an increase of the laser threshold by the excess gain $\Delta g_0 \equiv \Delta_{FM}^{2} / (2 \Delta_{AM})$ [see Eq.(7)]. In this way, a spectral window with net gain arises: cavity axial modes excited by noise at the center of the gainline can be transiently amplified, and then convected away from the center of the gainline because of frequency drift introduced by the FM modulation; Fig.1. Such a scenario of transient amplification of perturbations is analogous to the one predicted by K\"{a}rtner {\it et al.} in detuned AM ML \cite{r3}, however in the $\mathcal{PT}$-symmetric ML the drift dynamics occurs in the frequency (rather than time) domain and originates from asymmetric mode coupling. Like  in hydrodynamics models \cite{r36}, the dimensionless parameter $R$ that determines the amount of transient energy amplification plays the role of the Reynolds number \cite{r3}. Depending on the level of noise in the system, a sufficiently large value of $R$ can bring the system to a turbulent regime \cite{r3}. Like in Ref.\cite{r3}, in our model we did not consider spontaneous emission noise in the ML master equation (1), however just the noise  introduced in the numerical solution to Eq.(1) can induce a turbulence behavior for $R$ larger than $\simeq 27.6$, corresponding to a transient growth $G \sim 10^{24}$ \cite{r3}; turbulence can be observed at lower values of $R$, down to $R \sim 8$ \cite{r3}, if spontaneous emission noise is included in Eq.(1).  In the turbulent regime, the system does not reach a steady state, because it is non-periodically interrupted by a new spectrally-shifted pulse created out of the net gain spectral window that destroys the previous almost stationary pulse; see Fig.3(b) to be discussed below. 
While in the detuned AM ML the turbulent regime is always attained by increasing the detuning between modulation period and cavity round-trip time, in our $\mathcal{PT}$-symmetric ML turbulence can be prevented by the onset of $\mathcal{PT}$ symmetry breaking. In fact, the effective description of the $\mathcal{PT}$-symmetric ML in terms of the $\mathcal{PT}$-symmetric quantum oscillator Hamiltonian $\hat{H}_{eff}$ is accurate provided that $\Delta_{FM}$ remains smaller than $\Delta_{AM}$, i.e. in the unbroken $\mathcal{PT}$ phase. Therefore, the maximum energy transient growth that can be attained in the  $\mathcal{PT}$-symmetric ML can be estimated as $G_{max} \simeq \exp(2R_{max})$, where 
\begin{equation}
R_{max} \simeq (\omega_g / \omega_m) \sqrt{ \Delta_{AM}/2}
\end{equation}
 is obtained from Eq.(9) by assuming $\Delta_{FM}=\Delta_{AM}$ as an upper limit. In a typical ML laser $\Delta_{AM}$ is generally small ($\Delta_{AM} \sim 0.01-0.1$), therefore turbulence is observed when a sufficiently large number of cavity axial modes $N=\omega_g/ \omega_m$ falls within the laser gainline. In previous analysis, we neglected group velocity dispersion (GVD) effects, which is a reasonable assumption for ML pulses with duration down to a few ps, such as in ML Nd:YAG lasers. GVD would make $\mathcal{D}_g$ complex, thus breaking exact $\mathcal{PT}$ invariance of $\hat{H}$. However, even in the presence of small-to-moderate GVD, i.e. for $|{\rm Im}(\mathcal{D}_g)| \ll  {\rm Re}(\mathcal{D}_g)$, pulse splitting and turbulence discussed above can be still observed.\par 
 
 We checked the predictions of the theoretical analysis by direct numerical simulations of the ML laser equations (1) and (2) using a standard pseudo spectral split-step method and assuming a small random amplitude of the intracavity field $\psi(t,0)$ at initial round trip. Parameter values used in the simulations are $\gamma_{\|}=1 \times 10^{-3}$, $l=0.04$, $\Delta_{AM}=0.01$, and $g_0=0.15$, which are typical of solid-state lasers (e.g. Nd:YAG). Two different values of $N=\omega_g / \omega_m$ are considered. In an experiment, for a given modulation frequency $N$ can be controlled by changing the effective gain bandwidth $\omega_g$ using an intracavity etalon. In the first set of simulations, a relatively small value  $N=50$ is considered, and the ML pulse build-up dynamics was numerically simulated for increasing values of the FM amplitude $\Delta_{FM}$, from below to above the $\mathcal{PT}$ symmetry breaking transition. The results are shown in Fig.2. According to the theoretical predictions, below the symmetry breaking point a single ML pulse, with a spectrum shifted from the center of the gainline owing to the FM signal, is observed [Figs.2(a) and (b)], whereas pulse splitting is observed in the broken $\mathcal{PT}$ phase [Fig.2(c)]. In the latter case the amplitudes of the two ML pulses are generally different, depending on the initial random conditions. Note that, after an initial transient pulse built-up interval associated to relaxation oscillations, the highest peak pulse intensity shows a small but visible undamped oscillations in the broken $\mathcal{PT}$ phase [see the inset in Fig.2(c)]. Such oscillations arise from the interference of the two non-orthogonal and temporally-shifted  ML pulses  \cite{r19,r33}. Note also that the symmetry breaking transition does not show a turbulent behavior: indeed, for such a relatively narrow gain bandwidth the Reyonold number at the symmetry breaking point is $R_{max} \sim 3.53$  according to Eq.(11), which is smaller than the critical Reynolds number that brings the system to turbulence. On the other hand, turbulence can be observed by increasing the gain bandwidth. Figure 3 shows, as an example, numerical results obtained for $N=400$, corresponding to a Reynolds number $R_{max} \sim 28.3$ at the symmetry breaking point. Note that $\mathcal{PT}$ symmetry breaking is now associated to a turbulent behavior, with transiently growing spectral modes [indicated by the arrows in Fig.3(b)] that irregularly disrupt the stationary ML state.\par 
 In conclusion, the concept of $\mathcal{PT}$-symmetric ML has been introduced, in which asymmetric mode coupling  leads to a transition from a single to  double ML pulse emission. Such a transition is the signature of $\mathcal{PT}$ symmetry breaking and can show a turbulent behavior. The present results provide an important link between a fundamental operational regime of a laser, i.e. mode locking, and the emerging field of $\mathcal{PT}$ optics, suggesting that laser ML could provide a fertile laboratory tool to investigate the physics of $\mathcal{PT}$ symmetry in optics.

%%%%%%%%%%%%%%%%%%%%%%%%%%%%%%%
% References with full titles %
%%%%%%%%%%%%%%%%%%%%%%%%%%%%%%%
\newpage
%\footnotesize
 {\bf References with full titles}\\
 \\
 \noindent
1. A. E. Siegman, {\it Lasers} (University Science Books, Sausalito,
CA, 1986), Chap. 27.\\
2. H.A. Haus, {\it Mode-locking of lasers}, IEEE J. Sel. Top. Quantum Electron. {\bf 6},
1173 (2000).\\
3. U. Morgner and F. Mitschke, {\it Drift instabilities in the pulses from cw mode-locked lasers}, Phys. Rev. E {\bf 58}, 187 (1998).\\
4. F.X. K\"{a}rtner, D.M. Zumb\"{u}hl, and N. Matuschek, {\it Turbulence in mode-locked lasers}, Phys. Rev.
Lett. {\bf 82}, 4428 (1999).\\
5. S. Longhi and P. Laporta, {\it Excess noise in intracavity laser frequency modulation}, Phys. Rev. E {\bf 61}, R989 (2000).\\
6. B. Fischer, B. Vodonos, S. Atkins, and A. Bekker, {\it Experimental demonstration of localization in the frequency domain of mode-locked lasers with dispersion}, Opt. Lett. {\bf 27}, 1061 (2002).\\
7. A. Gordon and B. Fischer, {\it Phase Transition Theory of Many-Mode Ordering and Pulse Formation in Lasers}, 
Phys. Rev. Lett. {\bf 89}, 103901 (2002).\\
8. S. Longhi, {\it Dynamic localization and Bloch oscillations in the spectrum of a frequency mode-locked laser}, Opt. Lett. {\bf 30}, 786 (2005).\\
9. S. Yang and X. Bao, {\it Experimental observation of excess noise in a detuned phase-modulation harmonic mode-locking laser}, Phys. Rev. A {\bf 74}, 033805 (2006).\\
10. S. Longhi, {\it Metal-insulator transition in the spectrum of a frequency-modulation mode-locked laser}, Phys. Rev. A {\bf 77}, 015807 (2008).\\
11. J.B. Geddes, W.J. Firth, and K. Black, {\it Pulse Dynamics in an Actively Mode-Locked Laser}, SIAM J. Appl. Dyn. Syst. {\bf 2}, 647 (2003).\\ 
% G.H.C. New, M. Noy, J.A. Crosse, A. Rumley, L. Newson, Z.-Y. Chen, C. Cheung, and A. Todhunter, {\it Simulation of turbulence in mode-locked lasers}, Opt. Commun. {\bf 282}, 4418 (2009).\\
%12. R. Weill, B. Fischer, and O. Gat, {\it Light-Mode Condensation in Actively-Mode-Locked Lasers}, Phys. Rev. Lett. {\bf 104}, 173901 (2010).\\
12.  A. Rosen, R. Weill, B. Levit, V. Smulakovsky, A. Bekker, and B. Fischer, {\it Experimental Observation of Critical Phenomena in a Laser Light System},
Phys. Rev. Lett. {\bf 105}, 013905 (2010).\\
13. M. Leonetti, C. Conti, and C. Lopez, {\it The mode-locking transition of random lasers}, Nature Photon. {\bf 5}, 615 (2011).\\
%14. C. Lecaplain, J. M. Soto-Crespo, Ph. Grelu, and C. Conti, {\it Dissipative shock waves in all-normal-dispersion mode-locked fiber lasers}, Opt. Lett. {\bf 39}, 263 (2014).\\
%16.  D.V. Churkin, S. Sugavanam, N. Tarasov, S. Khorev, S.V. Smirnov, S.M. Kobtsev, and S.K. Turitsyn, {\it Stochasticity, periodicity and localized light
%structures in partially mode-locked fibre lasers}, Nature Commun. {\bf 6}, 7004 (2015).\\
14. S.-Y. Wu, W.-W. Hsiang, and Y. Lai, {\it Synchronous-asynchronous laser mode-locking transition}, Phys. Rev. A {\bf 92}, 013848 (2015).\\ 
15. G. Herink, B. Jalali, C. Ropers, and D. R. Solli, {\it Resolving the build-up of femtosecond mode-locking with single-shot spectroscopy at 90 MHz frame rate}, Nature Photon. {\bf 10}, 321 (2016).\\
16. K. G. Makris, L. Ge, and H. E. T\"{u}reci, {\it Anomalous Transient Amplification of Waves in Non-normal Photonic Media}, Phys. Rev. X {\bf 4}, 041044 (2014).\\
17. S. Longhi, {\it Phase transitions in Wick-rotated $\mathcal{PT}$-symmetric optics}, Ann. Phys. {\bf 360}, 150 (2015).\\ 
18. C.M. Bender, {\it Making sense of non-Hermitian Hamiltonians}, Rep. Prog. Phys. {\bf 70}, 947 (2007).\\
19. A Ruschhaupt, F Delgado, and J.G. Muga, {\it Physical realization of $\mathcal{PT}$-symmetric potential scattering in a planar slab waveguide}, J. Phys. A {\bf 38}, L171 (2005).\\
20. R. El-Ganainy, K.G. Makris, D.N. Christodoulides, and Z.H. Musslimani, {\it Theory of coupled optical $\mathcal{PT}$-symmetric structures}, Opt. Lett. {\bf 32}, 2632  (2007).\\
21. K. G. Makris, R. El-Ganainy, D. N. Christodoulides, and Z. H. Musslimani, {\it Beam Dynamics in $\mathcal{PT}$-Symmetric Optical Lattices}, Phys. Rev. Lett. {\bf 100}, 103904 (2008).\\
22. S. Longhi, {\it Bloch Oscillations in Complex Crystals with $\mathcal{PT}$ Symmetry}, Phys. Rev. Lett. {\bf 103}, 123601 (2009).\\
23. C. E. R\"{u}ter, K.G. Makris, R. El-Ganainy, D.N. Christodoulides, M. Segev, and D. Kip, {\it Observation of parity-time symmetry in optics}, Nature Phys. {\bf 6}, 192 (2010).\\
24. A. Regensburger, C. Bersch, M.-A. Miri, G. Onishchukov, D.N. Christodoulides, and U. Peschel, {\it Parity-time synthetic photonic lattices}, Nature {\bf 488}, 167 (2012).\\
25. L. Feng, Y.-L. Xu, W.S. Fegadolli, M.-H. Lu, J.E.B. Oliveira, V.R. Almeida, Y.-F. Chen, and A. Scherer, {\it Experimental demonstration of a unidirectional reflectionless parity-time metamaterial at optical frequencies}, Nature Mat. {\bf 12}, 108 (2013).\\
26.  S.V. Suchkov, A.A. Sukhorukov, J. Huang, S.V. Dmitriev, C. Lee, and Yuri S. Kivshar, {\it Nonlinear switching and solitons in PT-symmetric photonic systems}, Laser \& Photon. Rev. {\bf 10}, 177 (2016).\\
27. M. Kulishov, J.M. Laniel, N. Belanger, J. Azana, and D.V. Plant, {\it Nonreciprocal waveguide Bragg gratings }, Opt. Express {\bf 13}, 3068 (2005).\\
28. Z. Lin, H. Ramezani, T. Eichelkraut, T. Kottos, H. Cao, and D.N. Christodoulides, {\it Unidirectional Invisibility Induced by $\mathcal{PT}$-Symmetric Periodic Structures}, Phys. Rev. Lett. {\bf 106}, 213901 (2011).\\
29.  S. Longhi, {\it Invisibility in $\mathcal{PT}$-symmetric complex crystals}, J. Phys. A {\bf 44}, 485302 (2011).\\
30. S. Longhi, {\it Talbot self-imaging in $\mathcal{PT}$-symmetric complex crystals}, Phys. Rev. A {\bf 90}, 043827 (2014).\\
31. A.M. Dunlop, W. J. Firth, and E.M. Wright, {\it Time-domain master equation for pulse evolution and laser mode-locking}, Opt. Quant. Electron. {\bf 32}, 1131 (2000).\\
32. S. Longhi, {\it Pulse dynamics in actively mode-locked lasers with frequency shifting}, Phys. Rev. E {\bf 66}, 056607 (2002).\\
33. M.V. Berry, {\it Lop-sided diffraction by absorbing crystals}, J. Phys. A {\bf 31}, 3493 (1998).\\
34.  D.J. Kuizenga and A.E. Siegman, {\it FM and AM Mode Locking of the Homogeneous Laser - Part I: Theory}, IEEE J. Quantum Electron. {\bf 6}, 694 (1970).\\
35.  M Znojil, {\it $\mathcal{PT}$-symmetric harmonic oscillators}, Phys. Lett. A {\bf 259}, 220 (1999).\\
36. L. Trefethen, A. Trefethen, S. C. Reddy, and T. Driscol, {\it Hydrodynamic stability without eigenvalues},  Science {\bf 261}, 578 (1993).\\

 %F. Rana, R.J. Ram, and H.A. Haus, {\it Quantum Noise of Actively Mode-Locked Lasers
%With Dispersion and Amplitude/Phase Modulation}, IEEE J. Quant. Electron. {\bf 40}, 41 (2004).

\end{document}